\documentclass[aps,prl,preprint,superscriptaddress]{revtex4-1}

\usepackage{natbib}
\usepackage{amsmath}
\usepackage{graphicx}
\usepackage{nameref}
\usepackage{xcolor}
\usepackage{hyperref}
\usepackage{cleveref}

\begin{document}


\title{Nonadiabatic escape and stochastic resonance}


\author{W. Moon}
\email[]{woosok.moon@su.se}
\affiliation{Department of Mathematics, Stockholm University 106 91 Stockholm, Sweden}
\affiliation{Nordita, Royal Institute of Technology and Stockholm University, Stockholm 106 91, Sweden}
\author{N. Balmforth}
\email[]{njb@math.ubc.ca}
\affiliation{Department of Mathematics, University of British Columbia, Vancouver, BC V6T 1Z2, Canada}

\author{J. S. Wettlaufer}
\email[]{john.wettlaufer@yale.edu}
\affiliation{Yale University, New Haven, Connecticut 06520, USA}
\affiliation{Mathematical Institute, University of Oxford, Oxford OX2 6GG, UK}
\affiliation{Nordita, Royal Institute of Technology and Stockholm University, Stockholm 106 91, Sweden}



\date{\today}

\begin{abstract}
We analyze the fluctuation-driven escape of particles from a metastable state under the influence of a weak periodic force.  We develop an asymptotic method to solve the appropriate Fokker-Planck equation with mixed natural and absorbing boundary conditions. The approach uses two boundary layers flanking an interior region; most of the probability is concentrated within the boundary layer near the metastable point of the potential and particles transit the interior region before exiting 
the domain through the other boundary layer, which is near the unstable maximal point of the potential.   The dominant processes in each region are given by approximate 
 time-dependent solutions matched to construct the approximate composite solution, which gives the rate of escape with weak periodic forcing.  Using reflection we extend the method to a double well potential influenced by white noise and weak periodic forcing, and thereby derive a two-state stochastic model--the simplest treatment of stochastic resonance theory--in the nonadiabatic limit.   
\end{abstract}

\pacs{}

\maketitle


\section{Introduction}

The escape of particles from a metastable state under the influence of noise is a classical problem 
in non-equilibrium statistical mechanics \cite{kramers1940}.
Calculating the rate of escape can be approached in a variety of ways \cite[see e.g.,][and Refs. therein]{hanggi1990, vanKampen, Bressloff:2017, Forgoston:2018}. 
An important extension of the original escape problem includes the influence of periodic forcing, with a phase that impacts the escape rate \cite{nicolis1982, jung1993}.
Of particular relevance here is the problem of {\em stochastic resonance}, wherein the combined effect of background noise and weak periodic forcing 
control the state of the system  \cite{benzi1981, Benzi1982, gammaitoni1998}.  Indeed, because of the compelling consequences of such resonances, there are many methods that have been developed to calculate the escape rate, ranging from eigenfunction expansions \cite{jung1989} to path-integrals \cite{smelyanskiy1999, lehmann2000}.   However, the 
simplest solution used to study the principle characteristics of stochastic resonance appeals to the approximation of the adiabatic limit \cite{mcnamara1989}. 

Recently, concepts of stochastic resonance have been utilized in numerous fields including sensory biology \cite[e.g.,][]{vazquez2017, itzcovich2017}, 
image processing \cite[e.g.,][]{singh2016, gupta2016}, signal detection and processing \cite[e.g.][]{han2016, lai2016}, 
and energy harvesting \cite[e.g.,][]{zhang2016, kim2018}.  The broad impact of stochastic resonance is often viewed as counter-intuitive because rather than background noise obscuring the detection of a weak signal, it leads instead to an enhancement of that signal.   
 
The canonical configuration of stochastic resonance focuses on particles in a double-well potential influenced by white noise and weak periodic 
forcing.   Although there are multiple time-scales involved in the dynamics, the principal interest concerns the time it takes for a particle to transition from 
one stable point in the potential to the other.  Hence, it is common to consider a two-state model using a master equation that describes the time-evolution of the probability density of two discrete states and their exchange rates \cite{mcnamara1989}.  Thus one uses the classical escape rate from one metastable point; when the rate is {\em independent} of the slowly varying {\em phase} of the periodic forcing, this is called the {\em adiabatic limit}.   However, considering the extent of the fields in which stochastic resonance plays a role, from climate to engineering to biology \cite{wiesenfeld1995},  it is of interest to go beyond the adiabatic limit. Here we address the {\em nonadiabatic} situation, in the two-state framework, to determine the escape rate when the phase of the periodic forcing does not vary slowly.  To achieve this we introduce an asymptotic method to obtain an explicit expression for the escape rate, and in so doing we show how to transform
the original double-well potential problem into the two-state model.

\section{Escape rate under periodic forcing}\label{sec:escape_rate} 

\begin{figure}[ht]
\centering
\includegraphics[angle=0,scale=0.35,trim= 0mm 0mm 0mm 0mm, clip]{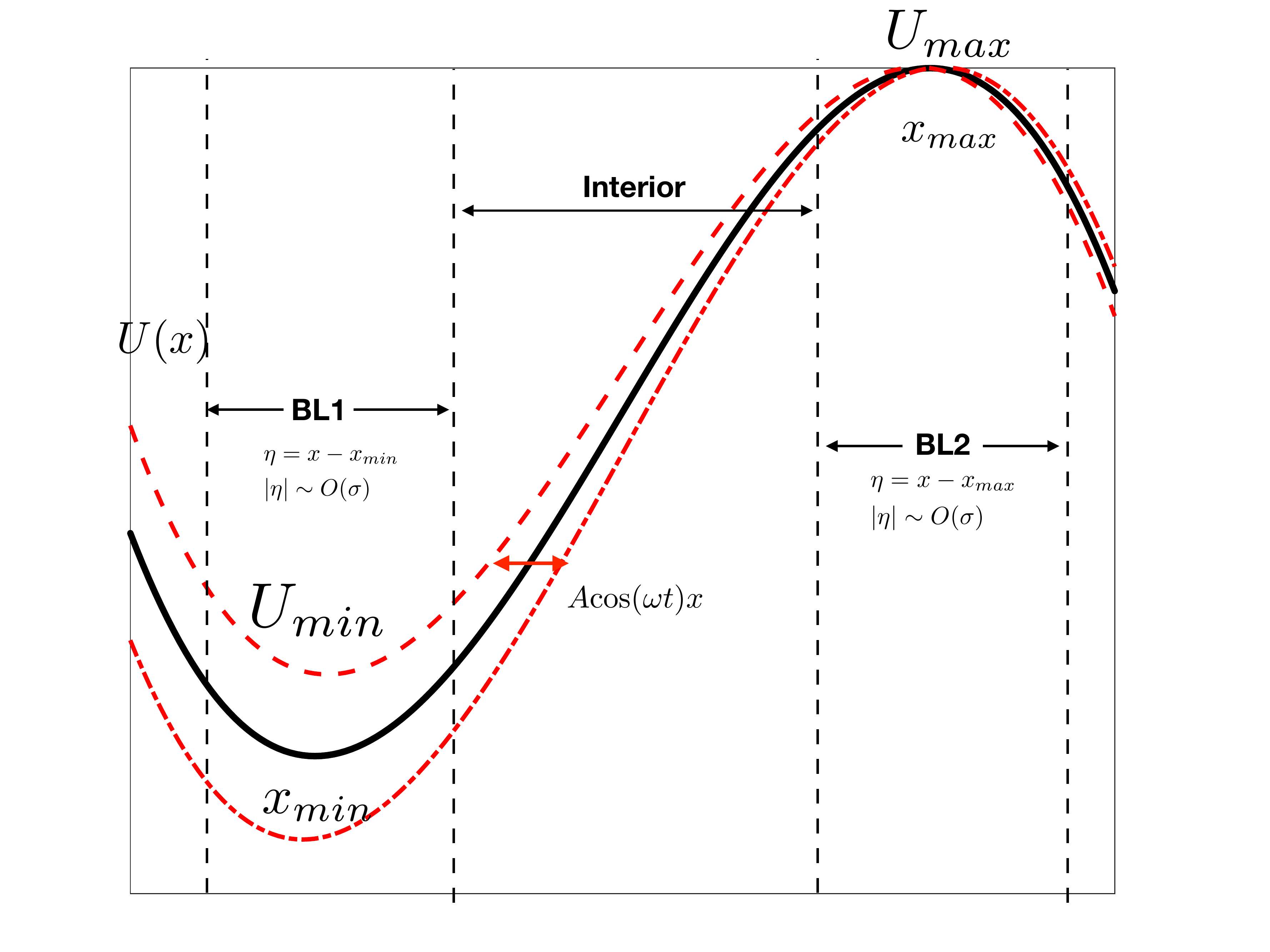}
\caption{Schematic of the escape rate problem with potential $U(x)$, periodic forcing $A\text{cos}(\omega t)$, and noise with magnitude $\sigma$. 
The width of the three regions {\bf BL1}, {\bf BL2} and {\bf Interior}, are overlain on the potential.}
\label{fig:schematic}
\end{figure}

First, we consider the escape rate from the metastable region of a potential $U(x)$ under the influence of
weak periodic forcing $A\text{cos}(\omega t)$ and noise induced fluctuations, as shown in Fig. \ref{fig:schematic}.  
The Fokker-Planck equation for this situation is
\begin{align}
   &\frac{\partial P}{\partial \tilde{t}} = \frac{\partial \tilde{J}}{\partial \tilde{x}} \qquad \textrm{where} \nonumber \\
   &\tilde{J}=\left[\frac{d\tilde{U}}{d\tilde{x}}-A\cos(\omega \tilde{t})\right]P+\sigma^2\frac{\partial P}{\partial \tilde{x}},
   \label{eq:FPE}
  \end{align}
with boundary conditions $P(\tilde{x}=-\infty,\tilde{t})=P(\tilde{x}=x_{max}+\chi,\tilde{t})=0$, wherein the tilde's denote dimensional variables.   
We assume that the magnitude of the noise, $\sigma$, and the amplitude of the
periodic forcing, $A$, are both small
(in the precise sense outlined below), and that $\chi=O(\Delta x)$.
The potential $U$ has the characteristic ``diffusivity'' scale,
$\Delta U = U_{max}-U_{min}$, and length scale, $\Delta x=x_{max}-x_{min}$, which
leads to the three parameters
\begin{align}
  \epsilon=\frac{\sigma}{\sqrt{\Delta U}}, \qquad
r=\frac{A\Delta x}{\sigma^2},
  \qquad \textrm{and}
  \qquad \Omega = \frac{\omega(\Delta x)^2}{\Delta U}. 
\end{align}
Using the dimensionless variables
 $x=(\tilde{x}-x_{max})/\Delta x$,
$t=\frac{\Delta U}{(\Delta x)^2}\tilde{t}$, $U=(\tilde{U}-U_{max})/\Delta U$
and $J=\tilde{J}/(\omega\Delta x)$,
Eq. \eqref{eq:FPE} becomes
\begin{align}
  \frac{\partial P}{\partial t}=
  \frac{\partial J}{\partial x}=
  \frac{\partial}{\partial x}
  \left[\frac{dU}{dx} - \epsilon^2
    r\cos(\Omega t)\right]P+\epsilon^2\frac{\partial^2 P}{\partial x^2} ,
  \label{eqn:main02}
\end{align}
with boundary conditions $P(x=-\infty,t)=P(x=\chi/\Delta x,t)=0$, and the local minimum and maximum are at $x=-1$ and $x=0$ respectively.

The underlying scaling assumptions are that
$\epsilon\ll1$ and  $r=O(1)$.
The small magnitude of $\epsilon$ is associated with the kinetic energy of a particle near the minimum of the potential being much less than the potential energy, $\Delta U$, necessary to escape from it.
  The assumption that the external forcing, $A\text{cos}(\omega t)$, is weak relative to the thermal noise is embodied by $r\sim O(\epsilon^2)$. The dimensionless frequency, $\Omega$, is the ratio of the time-scale for a particle to reach a quasi-stationary state near the potential minimum, $(\Delta x)^2/\Delta U$,
  to 
 the oscillation time-scale $\omega^{-1}$ of the potential.  The assumption that $\Omega \ll 1$ implies that the period of the external
forcing is much longer than the time required for a particle to reach a quasi-stationary state near the minimum. 
This assumption facilitates the asymptotic matching procedure near the maximum. 

\vspace{-0.5 cm}
\subsection{Potential Minimum -- Boundary Layer 1 (BL1): $|x+1| \sim O(\epsilon)$}
\vspace{-0.25 cm}

Near $x=-1$, the potential $U(x)$ can be approximated as
$U \simeq - 1 +\frac{1}{2}a(x+1)^2$, where $a \equiv |U''_{min}|$.
Thus, we rewrite Eq. \eqref{eqn:main02} in terms of a state variable $P_{B1}(\eta,t)$ that depends on the stretched coordinate
$\eta = (x+1)/\epsilon$:
\begin{align}
  \frac{\partial P_{B1}}{\partial t}=\frac{\partial}{\partial\eta}(a\eta P_{B1})
  - \epsilon r
  \cos(\Omega t)\frac{\partial P_{B1}}{\partial\eta}
+\frac{\partial^2P_{B1}}{\partial\eta^2}.
\end{align}
The leading-order solution,
written in terms of the original position variable, is
\begin{align}
P_{B1}(\eta,t)=n_1\sqrt\frac{{a}}{{2\pi\epsilon^2}}\text{exp}
\left[-\frac{a}{2\epsilon^2}(x+1)^2\right],
\label{eq:Pmin}
\end{align}
where $n_1$, which will be determined as part of the matching procedure, is a slowly-varying function of time satisfying $\frac{1}{n_1}\frac{dn_1}{dt} \ll 1$.
Implicit in this solution is therefore that the probability density
  reaches a quasi-steady state around the potential minimum, which arises
  because the weak noise in system drives only a small leakage of
  probability across the barrier at the maximum.

\vspace{-0.5 cm}
\subsection{Potential Maximum -- Boundary Layer 2 (BL2): $|x| \sim O(\epsilon)$} 
\vspace{-0.25 cm}

  Near the maximum ($x=x_{max}$) we let $x=\epsilon\zeta$ and the potential $U$
  can be approximated 
as $U \simeq -\frac{1}{2}b\epsilon^2\zeta^2$, where $b \equiv |U''_{max}|$.
Hence, Eq. \eqref{eqn:main02} is rewritten in terms of 
the state variable $P_{B2}(\zeta,t)$ as
\begin{align}
  \frac{\partial P_{B2}}{\partial t}=
  - \frac{\partial}{\partial\zeta}\left( b \zeta P_{B2}\right)
  +\frac{\partial^2 P_{B2}}{\partial\zeta^2} + O(\epsilon).
  \label{eq:max}
\end{align}
Equation \eqref{eq:max} must be solved subject to
$P_{B2}(\zeta=\infty,t)=0$ and that the solution match
to that for the interior region between the extrema of the potential,
outlined presently. The match, however, implies that the solution
for the interior delivers a probability flux to the boundary layer
around the maximum that varies periodically in time with frequency
$\Omega$. This precludes a straightforward solution
 of Eq. \eqref{eq:max} if $\Omega=O(1)$.

Instead, we avoid solving the boundary-layer problem for BL2 for general
$\Omega$, and adopt the convenient approximation that
the oscillation frequency is small,
$\Omega\ll1$. This allows us to neglect the left-hand side of Eq. \eqref{eq:max}
and write the quasi-stationary approximation,
\begin{align}
  P_{B2} \approx
  D_0\left[1-\sqrt{\frac{b}{2\pi\epsilon^2}} \int_{-\infty}^{x}
    \text{exp}\left(-\frac{bz^2}{2\epsilon^2}\right)dz\right]
  \text{exp}
\left(\frac{b x^2}{2\epsilon^2}\right).
\label{eq:pstarb2}
\end{align}

\vspace{-0.5 cm}
\subsection{Interior Region}
\vspace{-0.25 cm}

Within the interior region between the two extrema in the potential,
the rapid exponential decline of the probability $P$ suggests that 
we adopt a WKBJ-type ansatz, viz., 
\begin{align}
P \sim \text{exp}\left[-\epsilon^{-2}U+S(x,t)\right], 
\end{align}
from which the leading-order Fokker-Planck equation is
\begin{align}
  \frac{\partial S}{\partial t} + \frac{dU}{dx}\frac{\partial S}{\partial x} =
  r 
  \frac{dU}{dx}\text{cos}(\Omega t).
 \label{eq:S}
\end{align}
The characteristic curves of Eq. (\ref{eq:S}) are given by
\begin{align}
&\frac{dx}{dt} = \frac{dU}{dx} \equiv U'(x) \qquad \textrm{and}\nonumber \\
  &\frac{dS}{dx}= r 
  \text{cos}(\Omega t), 
\end{align}
which begin at $x=-1$ when $t\to-\infty$ where $S=S_{min}$, and converge to $x=0$ as $t\to\infty$ where $S=S_{max}$.
The solution is
\begin{align}
  S=S_{min}+ r 
  \int_{-1}^{x}\text{cos}\{\Omega\left[t+T(z)-T(x)\right]\}dz,
  \label{eq:ss}
\end{align}
where
\begin{align}
  T(x)=\int\frac{dx}{U'(x)}.
  \label{eq:T}
\end{align}
Note that, in the limit that $\Omega\ll1$, an integration by parts
furnishes the simpler approximation,
\begin{align}
  S &
  \simeq S_{min}+r(x+1)\text{cos}(\Omega t) 
  + r \Omega \sin(\Omega t) \int_{-1}^{x}(z+1)T'(z)dz,
 \label{eq:sx}
\end{align}
in which we set
  $\sin\{\Omega[t+T(z)-T(x)]\}\approx\sin\Omega t$,
assuming that $\Omega t$ may be $O(1)$ but $\Omega T(x)\ll1$.
This approximation exposes an issue with general solution in Eq. \eqref{eq:ss}: the transit time function $T(x)$ in Eq. \eqref{eq:T} diverges logarithmically for $x\to x_{min}$ or $x\to x_{max}$. This does not present a problem at the potential
minimum in view of the integration limits
in either Eq. (\ref{eq:ss}) or (\ref{eq:sx}),
but it does obscure the limit to the maximum. In fact,
  for $|x|=O(\delta)$ with $1\gg\delta\gg\epsilon$,
  the interior solution should actually
  be matched to the boundary-layer solution, leaving
$T \sim -b^{-1} \log\delta + O(1)$.
Thus we write Eq. \eqref{eq:T} as
\begin{align}
  S_{max} \sim S_{min}+ r 
  \int_{-1}^{\delta}\text{cos}\{\Omega\left[t+T(z)-T_{max}\right]\}dz,
  \label{eq:ssa}
\end{align}
where $T_{max}\sim t_0-b^{-1} \log \delta$ and
$t_0$ is an (undetermined) order-one constant time shift
that should, in principle, be fixed by a matching argument.

{\color{black}
\vspace{-0.5 cm}
\subsection{Asymptotic Matching \& Uniform approximation}
\vspace{-0.25 cm}

Asymptotic matching near the local minimum leads to
\begin{align}
  S_{min} \sim 
 \text{log}\left[P_{B1} e^{U/\epsilon^2 }   \right]_{\eta\to\infty}
\simeq \text{log}\left(n_1\sqrt{\frac{a}{2\pi\epsilon^2}}\right)
- \frac{1}{\epsilon^2}, 
\label{eq:s}
\end{align}
and near the local maximum it is required that
\begin{align}
  S_{max} \sim
\text{log}\left[P_{B2} e^{U/\epsilon^2 }   \right]_{\zeta\to-\infty}
\simeq \text{log}D_0 .
\label{eq:logD}
\end{align}

The preceding results suggest an approximation
that is valid throughout the two boundary layers and the interior
region:
\begin{align}
P \approx &
n_1\sqrt{\frac{a}{2\pi\epsilon^2}}
\left[1-\sqrt{\frac{b}{2\pi\epsilon^2}} \int_{-\infty}^{x}
    e^{-b z^2 / 2 \epsilon^2 } dz\right] \times \nonumber \\
    &\text{exp}\left[-\frac{(1+U)}{\epsilon^2} +
     r \int_{-1}^{{\rm min}(x,-\epsilon)}\text{cos}\{\Omega\left[t+T(z)-T(x)\right]\}dz\right].
     \label{uni_ori}
\end{align}
{\color{black}
In the small frequency ($\Omega \ll 1$) approximation, Eq. \eqref{uni_ori} becomes
  \begin{align}
 P \approx &
 n_1\sqrt{\frac{a}{2\pi\epsilon^2}}
   \left[1-\sqrt{\frac{b}{2\pi\epsilon^2}} \int_{-\infty}^{x}
    e^{-b z^2 / 2 \epsilon^2 } dz\right] \times \nonumber \\
   &\text{exp}\left[-\frac{(1+U)}{\epsilon^2} +
     r [{\rm min}(x,-\epsilon)+1]\text{cos}(\Omega t) +
  r \Omega \sin (\Omega t) \int_{-1}^{{\rm min}(x,-\epsilon)} (z+1) T'(z)dz\right].
   \label{unicorn}
  \end{align}
  }
Near $x=\pm1$, this approximation reduces to the two boundary layer solutions in Eqs. \eqref{eq:Pmin} and \eqref{eq:pstarb2},
with $D_0$ given by the relevant approximation of Eq. \eqref{eq:logD}, whereas in the interior it reduces to the solution implied by
Eq. \eqref{eq:sx}.
  Note that the limit of the last integral in (\ref{unicorn})
  introduces the approximation $\delta\approx\epsilon$, thereby furnishing a
  solution that depends on only a single small parameter and avoids
  any exercise in matching.

  
\vspace{-0.5 cm}
\subsection{Exit Rate}
\vspace{-0.25 cm}

Now, we construct the exit rate by a suitable integration of the Fokker-Planck
equation (\ref{eqn:main02}) as follows.  Note that
\begin{align}
\frac{d}{dt}\int_{-\infty}^{0}Pdx = \int_{-\infty}^{0}\frac{\partial J}{\partial x} dx = J|_{x=0},  
\label{eq:flux}
\end{align}
and because the probability is principally concentrated near the minimum, $x=-1$, we have 
\begin{align}
\int_{-\infty}^{0}Pdx \simeq \epsilon \int_{-\infty}^{\infty}P_{B1}d\eta = n_1.
\end{align}
The flux at the origin is given by 
\begin{align}
  J|_{x=0} \simeq \left. \epsilon^2\frac{\partial P}{\partial x}
  \right|_{x=0} 
 =-\epsilon^2D_0\sqrt{\frac{b}{2\pi\epsilon^2}}.
\end{align}
Hence, by combining Eqs. (\ref{eq:logD}) and (\ref{uni_ori}) we obtain
\begin{align}
  \frac{dn_1}{dt} = -\frac{\sqrt{ab}}{2\pi}
\text{exp}\left(-\frac{1}{\epsilon^2}+ r 
\int_{-1}^{\delta}\text{cos}\{\Omega[t+T(z)-T_{max}]\}dz\right)n_1, 
\label{eq:n1_evolution}
\end{align}
thereby giving the escape rate $R \equiv \frac{1}{n_1}\frac{dn_1}{dt}$ as
\begin{align}
 R &= \frac{\sqrt{ab}}{2\pi}
 \text{exp}\left( -\frac{1}{\epsilon^2} + r 
 \int_{-1}^{\delta}\text{cos}\{\Omega[t+T(z)-T_{max}]\}dz\right)
.
\label{eq:Rapprox}
\end{align} 
In the small frequency ($\Omega \ll 1$) approximation,
by substituting Eq. (\ref{unicorn}) into $J|_{x=0}=\epsilon^2\partial P / \partial x|_{x=0}$
we have
\begin{align}
 R =  \frac{\sqrt{ab}}{2\pi}
 \text{exp}\left(-\frac{1}{\epsilon^2}+r\text{cos}\Omega t+r\Omega\text{sin}\Omega t\int_{-1}^{\delta}(z+1)T'(z)dz\right),
  \label{eq:Rsmallomega}
\end{align} 
}


\vspace{-0.5 cm}
\subsection{Adiabatic Limit}
\vspace{-0.25 cm}

In the adiabatic limit, we discard the terms containing factors of $\Omega$
in Eq. \eqref{unicorn}, to arrive at
\begin{align}
 P_{ad} = 
 n_1\sqrt{\frac{a}{2\pi\epsilon^2}}
   \left[1-\sqrt{\frac{b}{2\pi\epsilon^2}} \int_{-\infty}^{x}
    e^{-b z^2 / 2 \epsilon^2 } dz\right] 
   \text{exp}\left[-\frac{(1+U)}{\epsilon^2} +
     r(x+1)\text{cos}(\Omega t) \right],
   \label{unicorna}
\end{align}
with the associated escape rate 
\begin{align}
 R_{ad} & =  
 \frac{\sqrt{ab}}{2\pi}\text{exp}\left[
 r \cos(\Omega t) - \frac{1}{\epsilon^2}\right].
 \label{eq:Rad}
\end{align} 

\vspace{-0.5 cm}
\subsection{The cubic potential}
\vspace{-0.25 cm}

For the cubic potential 
\begin{align}
  \tilde{U} = U_{max}-\Delta U \left[
    \frac{(\tilde{x}-x_{max})^2}{2(\Delta x)^2}+
    \frac{(\tilde{x}-x_{max})^3}{3(\Delta x)^3}\right],
\end{align}
we have
\begin{equation}
  U=-3x^2-2x^3, \qquad
  \frac{dU}{dx}=-6x(x+1),  \qquad {\rm and} \qquad
  a=b=6.
\end{equation}
Hence we have
\begin{align}
 T(x) = \frac{1}{6}\text{log}\left(\frac{1+x}{-x}\right),
\end{align}
\begin{align}
  S=S_{min}+ \frac{3r}{2} 
  \int_{-\infty}^{T(x)}\frac{\text{cos}\{\Omega[\tau+t-T(x)]\}}
      {\text{cosh}^2(3\tau)}d\tau, 
\end{align}
and
\begin{align}
  R =
  \frac{3}{\pi}\exp\left[-\frac{1}{\epsilon^2} +
    r \Upsilon(\Omega) \cos\theta\right],
 \label{eq:nonadR}
\end{align}
where
$$
\theta \equiv \Omega(t-T_{max}) \simeq \Omega \left( t+
\frac{1}{6}\log\epsilon \right),
$$
if we again set $\delta=-\epsilon$, and
\begin{equation}
  \Upsilon(\Omega) = \frac{\frac{1}{6}\Omega\pi}{\sinh(\frac{1}{6}\Omega\pi)}
  \label{Uppy}
\end{equation}
  captures the suppression of the periodic adiabatic variation
  of the escape rate by non-adiabatic effects. 

\begin{figure}[ht]
\centering
\includegraphics[angle=0,scale=0.45,trim= 0mm 20mm 0mm 30mm, clip]{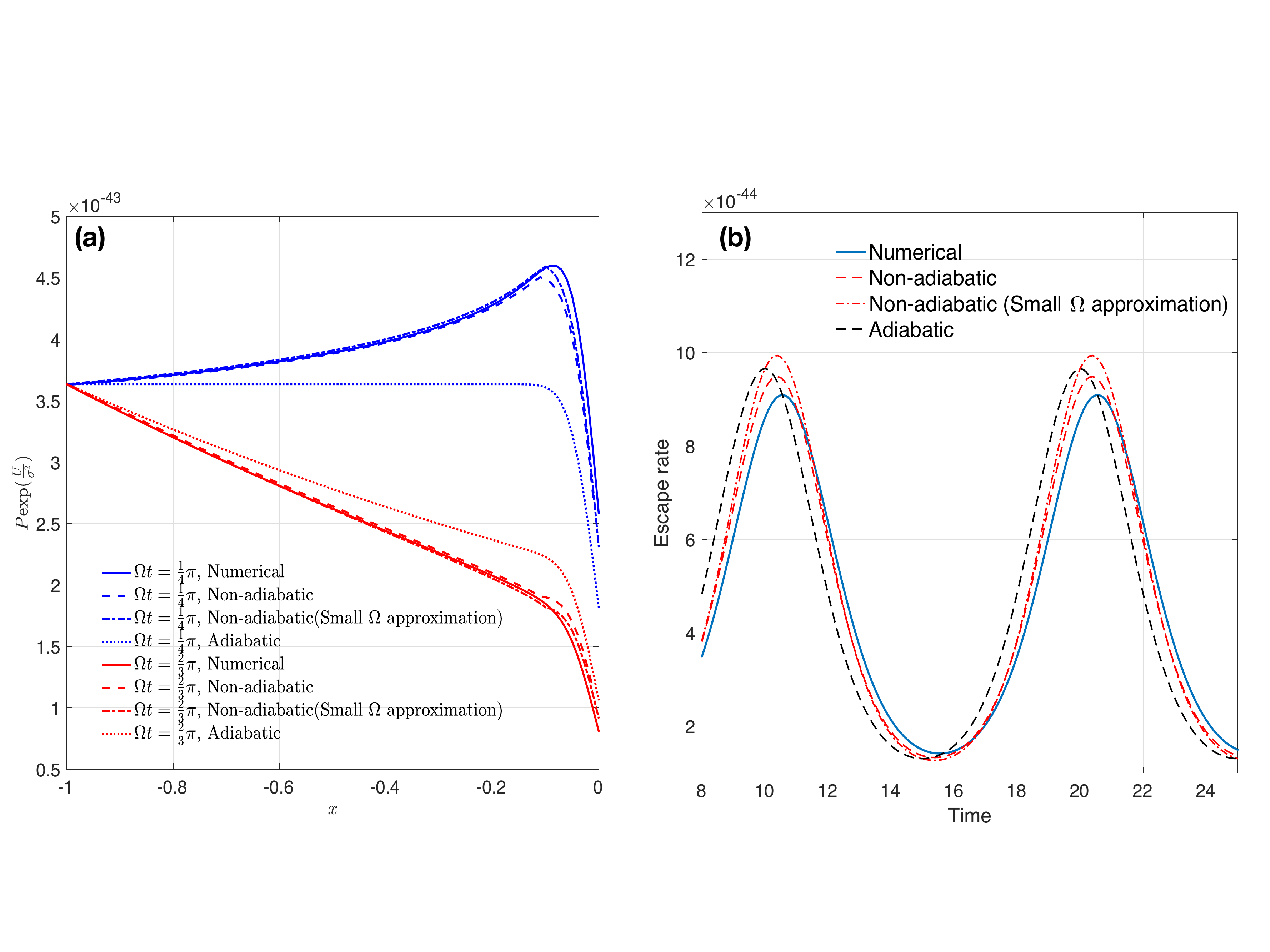}
\caption{ {\color{black} (a) 
Comparison  of the numerical solution to the Fokker-Planck Equation (Eq. \ref{eqn:main02}, solid), with 
the non-adiabatic (Eq. \ref{uni_ori}, dashed), 
the non-adiabatic in the small $\Omega$ approximation (Eq. \ref{unicorn}, dashed-dot) and the adiabatic (Eq. \ref{unicorna}, dotted) analytic solutions, in the case of the cubic potential $U(x)=-3x^2-2x^3$.  
The parameter values are
{\color{black} $r=1$, $\epsilon=0.1$ and $\Omega = \pi/5$.}
(b) The associated escape rates, {\color{black}$J_{x=0}/\int_{-\infty}^0 P dx$},  for the numerical (solid), non-adiabatic (Eq. \ref{eq:nonadR}, dashed red), 
non-adiabatic in the small $\Omega$ approximation (Eq. \ref{eq:Rsmallomega}, dashed-dot)
and adiabatic (Eq. \ref{eq:Rad}, dashed black) analytic solutions. }}
\label{fig:smallomega}
\end{figure}
  
{\color{black}In Fig.~\ref{fig:smallomega}(a) we compare a numerical solution
of the Fokker-Planck Equation, \eqref{eqn:main02}, with
the non-adiabatic (Eqs. \ref{uni_ori} and \ref{unicorn}) and adiabatic (Eq. \ref{unicorna}) analytical solutions.  
Our numerical method for Eq. \eqref{eqn:main02} is based on the implicit finite difference scheme introduced by Chang and Cooper \cite{Chang}.
Both of the non-adiabatic analytical solutions match the numerical solution at the
percentage level of accuracy, save for the transition region from the interior to the boundary layer near the maximum ($x=0$). 
 However, the adiabatic solution differs substantially from both of the others, as is particularly evident when $\Omega t = \pi/2$ 
 where the non-adiabatic contribution, $\text{sin}(\Omega t)$, is maximal.   

In Fig.~\ref{fig:smallomega} (b) we show the time evolution of the escape rates for the four solutions, defined as $J_{x=0}/\int_{-\infty}^0 P dx$ for the numerical solution.
To calculate the analytical solutions, we use the approximation $n_1=1$, which is accurate
to machine precision for the parameter settings and times used in the figure. (Likewise, for the double-well potential below, we use the approximation $n_1=n_2=0.5$ in comparing the asymptotic predictions with numerics in Fig.~\ref{fig:sto_resonance}.) The non-adiabatic analytical
solutions compare well with the numerical solution, whereas there is a pronounced deviation of the adiabatic solution in both phase and amplitude of the maximum escape rates.

Finally, in Fig.~\ref{fig:esccomp} we bring out the deviations of the various approximations as a function of frequency.  Note in particular the substantial differences in phase and amplitude between Figs.~\ref{fig:esccomp}(a) and (c).  In particular, while
the non-adiabatic analytic solutions compare well with the numerical solution, there is a pronounced deviation of the adiabatic solution in both phase and amplitude of the maximum escape rates.  

\begin{figure}[ht]
\centering
\includegraphics[angle=0,scale=0.45,trim= 40mm 0mm 30mm 0mm, clip]{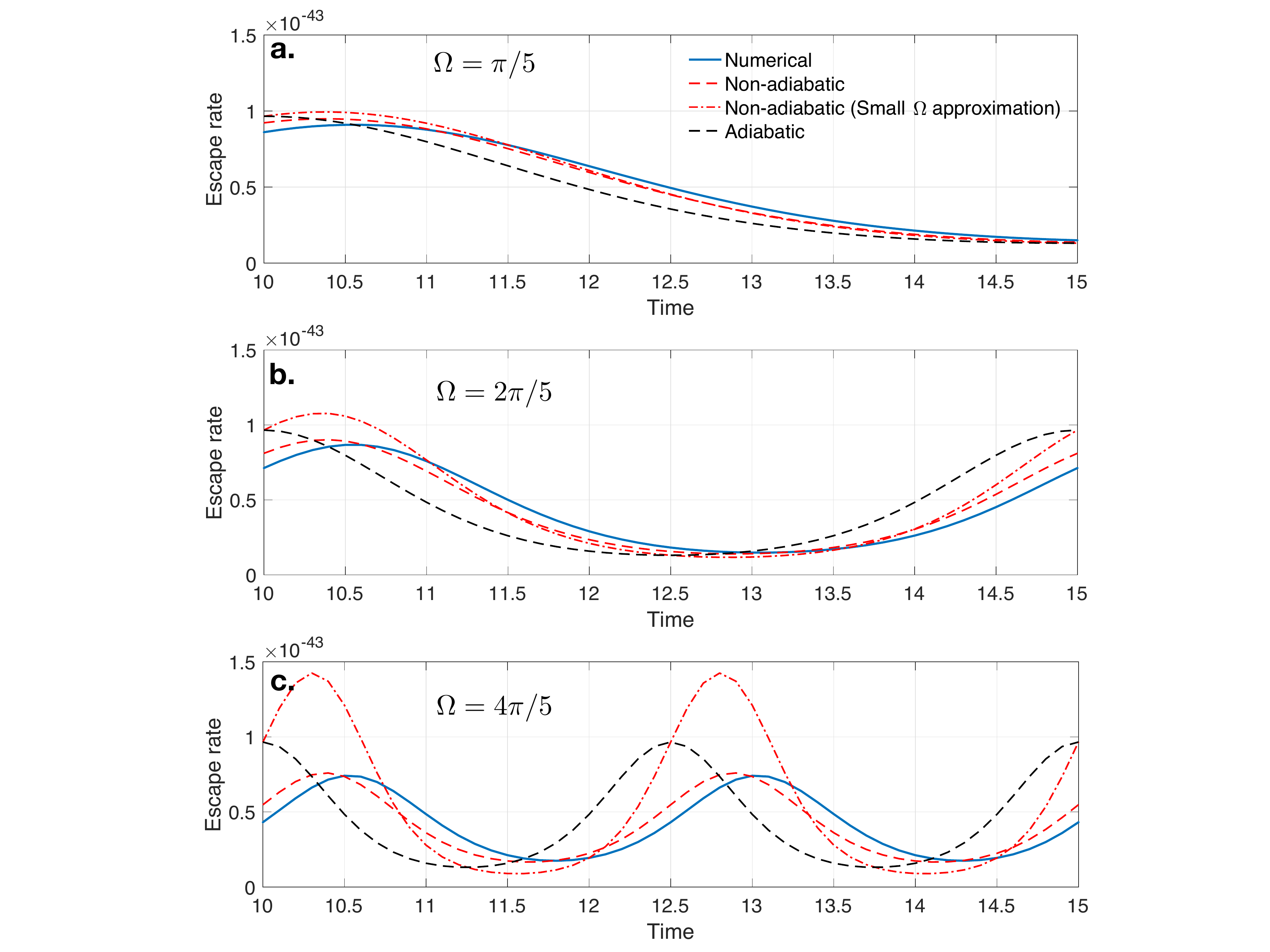}
\caption{
Comparison of the escape rate calculated from the numerical solution (solid blue), the non-adiabatic solution (dashed red),
the non-adiabatic small $\Omega$ approximate solution (dashed-dot red), and the adiabatic approximation (dashed black) for
$\Omega$ = 1/5$\pi$ (a), 2/5$\pi$ (b) and 4/5$\pi$ (c). The potential, $U(x)=-3x^2-2x^3$, and the parameter values, $r=1$ and $\epsilon=0.1$, 
are the same as in Fig.~ \ref{fig:smallomega}. 
}
\label{fig:esccomp}
\end{figure}


\section{Double-well potential and Stochastic Resonance}

We now treat Brownian particles in a double-well potential under the influence
of weak periodic forcing, which is the  
original configuration of stochastic resonance \cite{benzi1981, Benzi1982}.
By reflection of Fig. \ref{fig:schematic} we extend the approach described
above to construct the approximate solutions in the five regions
shown in Fig. \ref{fig:schematic02}.
The potential $\tilde{U}$ is scaled as before, so that
$U=(\tilde{U}-U_{max})/\Delta U$, where $\Delta U$ is now a measure of the
height of the barrier, and we define $\Delta x$ as half the distance
between the two minima. As the potential may not be symmetrical,
this translates to a scaled potential that vanishes at $x=0$
and takes the values $U_{1}$ and $U_{2}$ at the two minima
$x_{1}$ and $x_{2}$, respectively. 
We replace the absorbing boundary near the 
local maximum with the usual boundary condition, $P(\pm\infty, t)=0$,
insuring that the probability is conserved throughout 
the entire domain as particles move between the two minima.

\begin{figure}[h]
\centering
\includegraphics[angle=0,scale=0.35,trim= 0mm 0mm 0mm 0mm, clip]{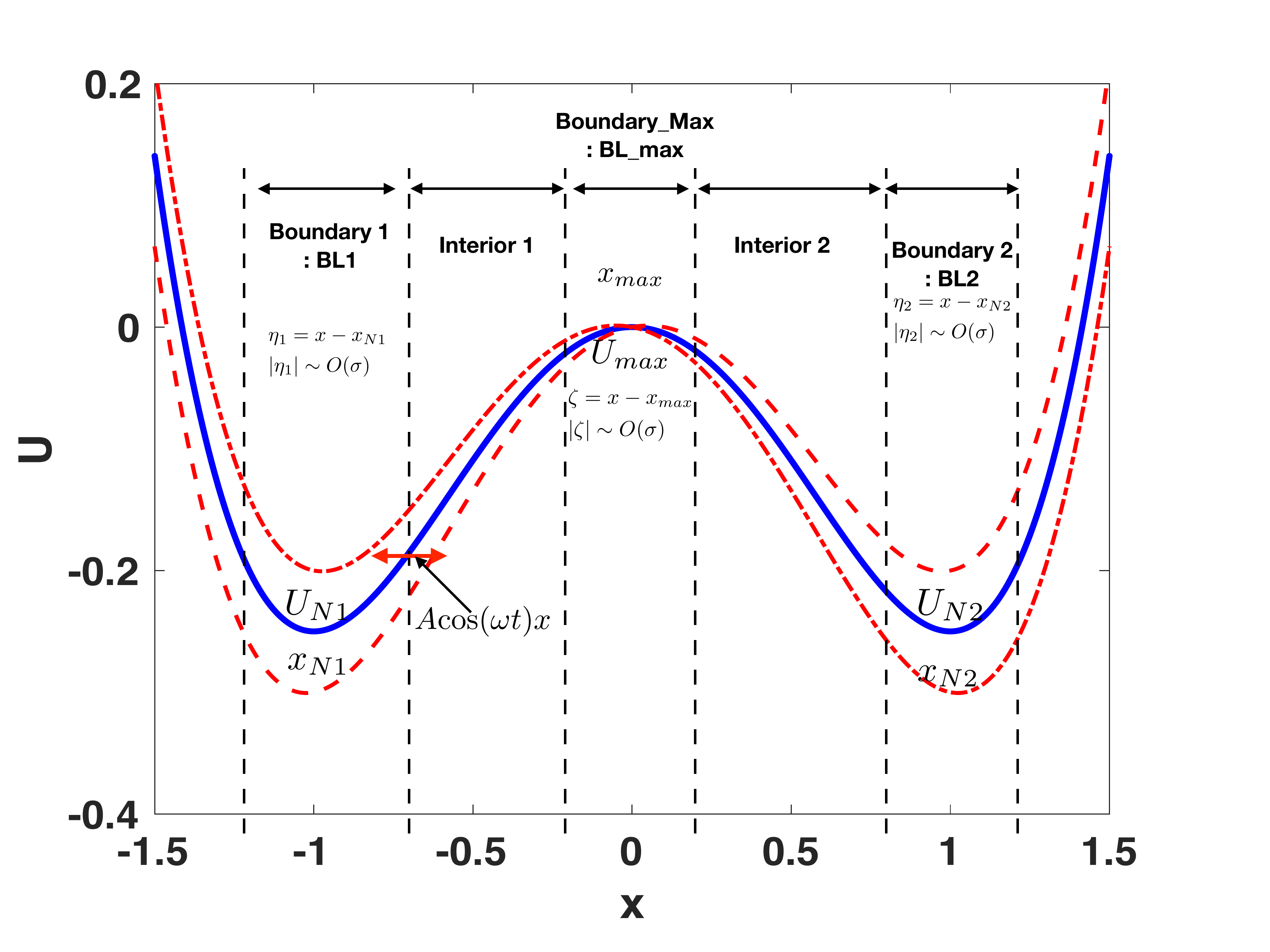}
\caption{Schematic of the five regions in the double-well potential $U(x)$ under the influence of weak periodic forcing, $A\text{cos}(\omega t)$, in which we find approximate solutions to the Fokker-Planck Equation. }
\label{fig:schematic02}
\end{figure} 
\vspace{-0.25 cm}

The asymptotic solution is
\begin{align}
P = 
\begin{cases}
  n_1 \sqrt{\frac{a_1}{2\pi\epsilon^2}}
  \exp \left[ - \frac{a_1}{2\epsilon^2}(x-x_1)^2\right],
  &  { {\rm BL}_{1}} \nonumber\\
  \exp\left(S_1-\frac{U}{\epsilon^2}
  +r\int_{-1}^{x}\cos[\Omega(t+T(z)-T(x))]dz\right),
  &  { {\rm Interior} \ 1} \nonumber \\
  \left[ D_0+D_1\sqrt{\frac{b}{2\pi\epsilon^2}}\int_{-\infty}^{x}
    \exp\left(-\frac{bz^2}{2\epsilon^2}\right)dz \right]
  \exp \left(  \frac{bx^2}{2\epsilon^2}\right),
  &  { {\rm BL}_\text{max}} \nonumber\\
  \exp\left(S_2-\frac{U}{\epsilon^2}
  +r\int_{1}^{x}\cos[\Omega(t+T(z)-T(x))]dz\right),
  & { {\rm Interior} \ 2} \\
  n_2 \sqrt{\frac{a_2}{2\pi\epsilon^2}}
  \exp \left[ - \frac{a_2}{2\epsilon^2}(x-x_2)^2\right],
  &  { {\rm BL}_{2}} \nonumber\\
 \end{cases} 
 \label{eq:sol_double}                     
\end{align}
where $a_{\{1,2\}}=U''(x_{\{1,2\}})$, and the ``constants'' of integration,
$n_1$, $n_2$, $S_1$, $S_2$, $D_0$ and $D_1$
must be connected by matching the five solutions together. In particular,
we find
\begin{align}
 S_{\{1,2\}} = \log\left(n_{\{1,2\}}\sqrt{\frac{a_{\{1,2\}}}{2\pi\epsilon^2}}\right)
  +\frac{U_{\{1,2\}}}{\epsilon^2},
\end{align}
\begin{align}
  &D_0=\text{exp}\left(S_1+r
  \int_{-1}^{-\epsilon}\cos[\Omega(t+T(x)-T(-\epsilon))]dx
  \right) \qquad \text{and} \nonumber \\
  &D_0+D_1=\text{exp}\left(S_2+r
  \int_{1}^{\epsilon}\cos[\Omega(t+T(x)-T(\epsilon))]dx\right),
\end{align}
if we again make the approximation that the match can be accomplished at
$x=\pm\epsilon$.
 Again, more compactly we have 
 \begin{align}
  P &\simeq n_1\sqrt\frac{{a_1}}{{2\pi\epsilon^2}}\text{exp}\left(-\frac{U}{\epsilon^2}
  +r\int_{-1}^{\text{min}(x,-\epsilon)}\text{cos}[\Omega(t+T(x')-T(\text{min}(x,-\epsilon)))]dx'\right) \nonumber \\
  &+\left[n_2\frac{\sqrt{a_2 b}}{2\pi\epsilon^2}\text{exp}\left(-\frac{U}{\epsilon^2}
  +r\int_{1}^{\text{max}(x,\epsilon)}\text{cos}[\Omega(t+T(x')-T(\text{max}(x,\epsilon)))]dx'\right) \right. \nonumber \\
  &\left. -n_1\frac{\sqrt{a_1 b}}{2\pi\epsilon^2}\text{exp}\left(-\frac{U}{\epsilon^2}
  +r\int_{-1}^{-\epsilon}\text{cos}[\Omega(t+T(x')-T(-\epsilon))]dx'\right) \right]
    \int_{-\infty}^{x}\text{exp}\left(-\frac{b}{2\epsilon^2}x'^2\right)dx'.
    \label{uni_ori_double}
 \end{align}
{\color{black} In the small frequency ($\Omega \ll 1$) approximation, Eq. \eqref{uni_ori_double} becomes
\begin{align}
 P &\simeq n_1\sqrt\frac{{a_1}}{{2\pi\epsilon^2}}\text{exp}\left(-\frac{U+1}{\epsilon^2}
                  +r(x+1)\text{cos}(\Omega t)+r\Omega\text{sin}(\Omega t)\int_{-1}^{\text{min}(x,-\epsilon)}
                  (z+1)T'(z)dz\right) \nonumber \\
     &+\left[n_2\frac{\sqrt{a_2 b}}{2\pi\epsilon^2}\text{exp}\left(-\frac{U+1}{\epsilon^2}
     +r(x-1)\text{cos}(\Omega t) + r\Omega\text{sin}(\Omega t)\int_{1}^{\text{max}(x,\epsilon)}(z-1)T'(z)dz\right) \right. \nonumber \\
     &\left. -n_1\frac{\sqrt{a_1 b}}{2\pi\epsilon^2}\text{exp}\left(-\frac{U+1}{\epsilon^2}
     +r(1-\epsilon)\text{cos}(\Omega t)+r\Omega\text{sin}(\Omega t)\int_{-1}^{-\epsilon}(z+1)T'(z)dz\right)\right]
      \int_{-\infty}^{x}\text{exp}\left(-\frac{b}{2\epsilon^2}x'^2\right)dx'.
      \label{uni_ori_smallomega}             
 \end{align}
}
Global conservation of probability implies that
$\int_{-\infty}^{\infty}P \simeq n_1+n_2 = 1$,  where 
$\frac{dn_1}{dt}=J|_{x=x_{max}}$, which leads to
\begin{align}
 \frac{dn_1}{dt} = \epsilon^2D_1\sqrt{\frac{b}{2\pi\epsilon^2}},
\end{align}
or equivalently
\begin{align}
  \frac{dn_1}{dt}=-R_1n_1+R_2n_2=R_2-(R_1+R_2)n_1,
  \label{SM}
\end{align}
where 
\begin{align}
  R_{\{1,2\}} =
  \frac{\sqrt{a_{\{1,2\}}b}}{2\pi}
  \exp\left(-\frac{U_{\{1,2\}}}{\epsilon^2}\right)
  \exp\left(r \Upsilon_{\{1,2\}}
  \cos [\Omega(t-T(\{-\epsilon,\epsilon\})) + \Theta_{\{1,2\}} ]
  \right)
  \label{escape2}
\end{align}
are the escape rates from $x_{\{1,2\}}$ through $x_{max}$,  with
\begin{equation}
  \int_{\{-1,1\}}^{\{-\epsilon,\epsilon\}}
  \cos[\Omega(t+T(x)-T(\{-\epsilon,\epsilon\}))]  dx 
  \equiv
  \Upsilon_{\{1,2\}}(\Omega)
  \cos [\Omega(t-T(\{-\epsilon,\epsilon\})) + \Theta_{\{1,2\}}(\Omega)].
\end{equation}
Equation \eqref{SM} has the same form as the
two-state Master equation used in other
approaches to stochastic resonance theory \cite[e.g.,][]{mcnamara1989}.
The factors $\Upsilon_{\{1,2\}}(\Omega)$ again represent the
suppression of the adiabatic variation of the escape rates by
non-adiabatic effects; $\Theta_{\{1,2\}}(\Omega)$ represent
additional phase shifts. The functions $\Upsilon_{\{1,2\}}(\Omega)$
are equal for a symmetrical base potential, and are
illustrated in figure \ref{fako}
for the quartic potential with $U=\frac{1}{4}x^4-\frac{1}{2}x^2$.
In the theory
of stochastic resonance presented in \cite{mcnamara1989},
the adiabatic variation of the escape rates (given
by the replacements $\Upsilon_{\{1,2\}}(\Omega)\to1$
and $\Theta_{\{1,2\}}(\Omega)\to0$ in (\ref{escape2}))
is responsible for the characteristic improvement in the signal-to-noise ratio.
Thus, the suppression factors  $\Upsilon_{\{1,2\}}(\Omega)$
determine the destructive effects of non-adiabaticity in a
generalization of that analysis.
In view of our small-frequency approximation, the exponential
decline of $\Upsilon_1=\Upsilon_2$ for large $\Omega$ seen in
figure \ref{fako} is again only approximate.

 \begin{figure}[ht]
\centering
\includegraphics[width=0.65\linewidth]{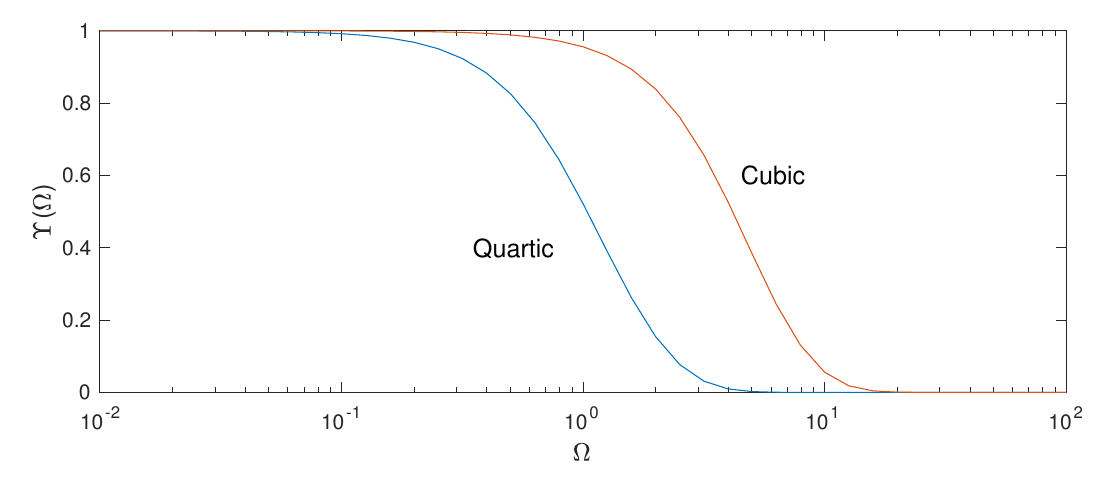}
\caption{ \label{fako}
  The function $\Upsilon_1=\Upsilon_2= \Upsilon(\Omega)$ for the quartic
  potential with $U=\frac{1}{4}x^4-\frac{1}{2}x^2$;
  also shown is the same function
  defined in (\ref{Uppy}) for the cubic potential.}
\end{figure}

In Fig.~\ref{fig:sto_resonance} we compare the numerical and analytic
solutions from Eqs. (\ref{uni_ori_double}) and (\ref{uni_ori_smallomega}), for the double well potential
$U=\frac{1}{4}x^4-\frac{1}{2}x^2$.
Because our approach is a reflection of
the escape calculation described in the previous section, we expect the match
to be good, save for the same transition region, here between the two interior regions and the central maximum.

\begin{figure}[ht]
\centering
\includegraphics[angle=0,scale=0.4,trim= 0mm 0mm 0mm 0mm, clip]{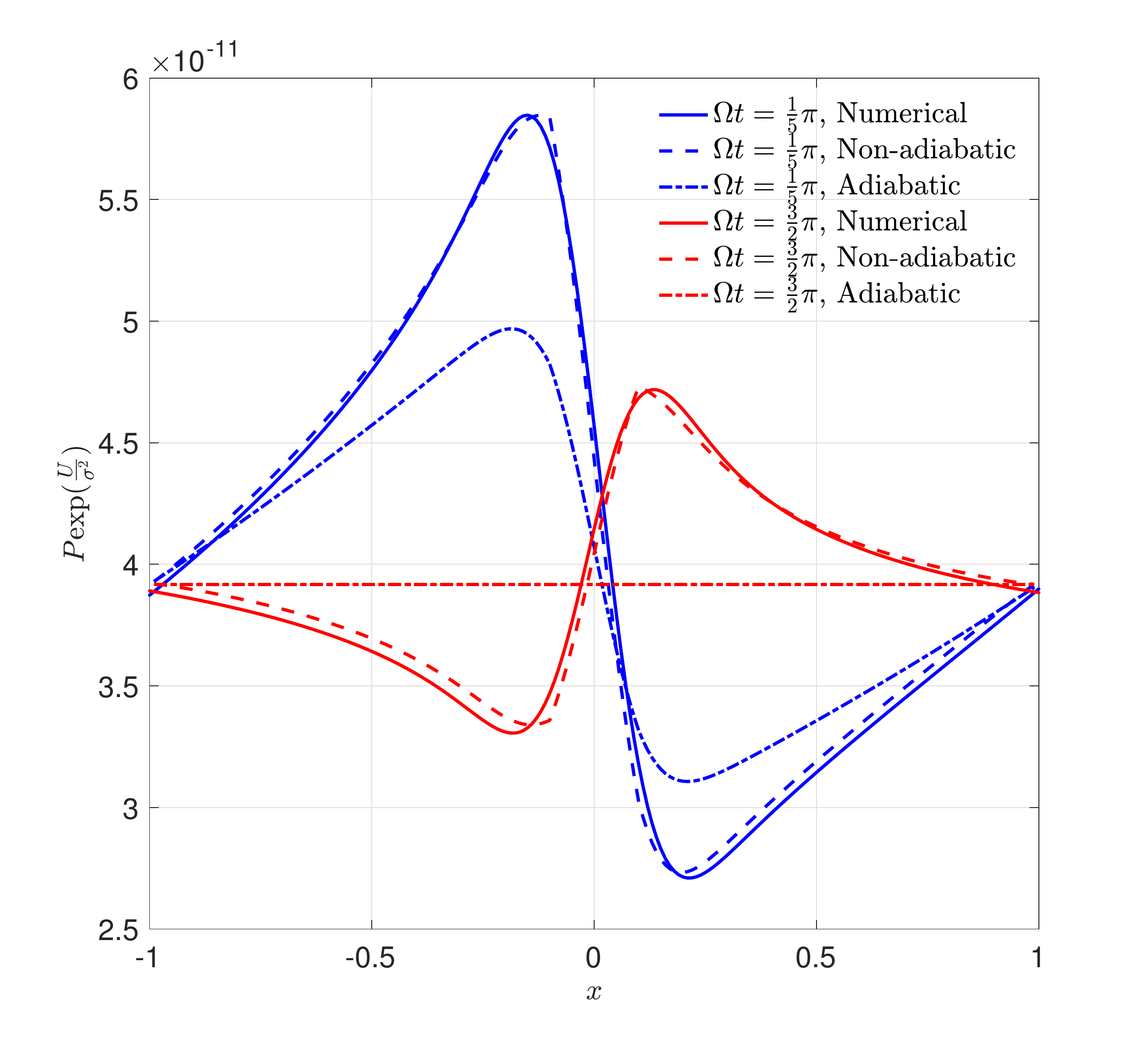}
\caption{{\color{black} We compare the numerical (solid), non-adiabatic (Eq. \ref{uni_ori_double}, dashed) and non-adiabatic in the small $\Omega$ approximation (Eq. \ref{uni_ori_smallomega}, dashed-dot) analytic solutions for the probability density profiles, 
$P\text{exp}\left(\frac{U}{\sigma^2}\right)$, in the double-well potential $U=\frac{1}{4}x^4-\frac{1}{2}x^2$ with  $A=0.01$, $\sigma=0.1$ and $\omega = \pi/20$, which is equivalent to $r=1$, $\epsilon=0.2$ and $\Omega=4\omega$}.}
\label{fig:sto_resonance}
\end{figure}

\section{Conclusion}

We have developed an asymptotic method of calculating the probability density function and the associated 
escape rate of Brownian particles from 
a metastable state under weak periodic forcing.  The approach uses boundary layers near the two extremes, where the potential $U(x)$ is approximately quadratic and the time-dependent linear Fokker-Planck equations can be solved. 
In the interior layer separating these, an advection-dominated solution is constructed and the three approximate solutions are matched.  Because the evolution of the total probability is equal to the probability flux at the absorbing boundary, we can integrate Fokker-Planck Equation over the complete domain and determine the escape rate in the non-adiabatic limit.  Finally, by reflection we extended this asymptotic approach 
to the problem of a double-well potential with weak periodic forcing to find a solution to the problem of stochastic resonance 
in the non-adiabatic case.  In particular, the ease with which Eq.~\ref{escape2} can be used, and its limits understood through Fig.~\ref{fako}, provide substantial applicability.  Given the ubiquity of stochastic resonance, this result is likely of the broadest relevance.  Additionally, the approach we take here is complimentary to other general approaches, which focus on the universality of fast-slow systems in stochastic resonance and two state systems \cite{BerglandEPL, BerglandBook, SoonHoe19}. 

\begin{acknowledgements}
WM and JSW acknowledge the support of Swedish Research Council grant no. 638-2013-9243.  
WM acknowledges a Herchel-Smith postdoctoral fellowship
and JSW a Royal Society Wolfson Research Merit Award for support. 
 \end{acknowledgements}


%

\end{document}